# Towards Implementing an Enterprise Groupware-Integrated Human Resources Information System

Sergey V. Zykov
IT Department
ITERA International Energy Corporation,
Moscow, Russia
szykov@itera.ru

**Abstract**

Human resources management software is having a wide audience at present. However, no truly integrate solution has been proposed yet to improve the systems concerned. Approaches to extra data collection for appraisal decision-making are considered on the concept modeling theoretical basis. Current technologies in state-of-the-art HR management software are compared. Design and implementation aspects for a Web-wired truly integrated secure and scalable event-driven enterprise system are described. Benchmark results are presented. Field perspectives are discussed.

## 1. Introduction

Frequent priority changes in versatile corporations development demand fast and flexible adaptation of personnel organizational structure to rapidly changing modern market with stiff competition. Such adaptation should be based on strategic software unification, and is especially urgent for human resources information and management systems (HRIMS) that involve a collection of technologies allowing to represent complex processes that center on personnel activities.

During the two last decades accents in data models, applied methods and software development tools have been shifted from imperative and operator-based to object-oriented programming. During this period information system (IS) architectures have evolved from mainframes to database and application servers capable of mobile and interoperable data processing. Narrow application spectrum IS based on early DBMS without any data model (DM) foundation have been substituted by more uniform file-server systems of versatile material resource planning (MRP) supporting E.F.Codd's relational DM. Since early 90-s the latter IS have been developed up to high-end well-consolidated enterprise resource planning (ERP) systems based on extendable relational DM, object-relational and object-oriented DBMS; attempts of enterprise application integration have also been undertaken.

Primary objective of the paper is analyzing, designing and implementing approaches to construct large enterprise software systems aimed at conglomerate resource accounting, planning and management. The approach suggested has been practically approved while prototyping and implementing full-scale HRIMS.

Major goals achieved can be summarized as follows:

- ERP software development approaches analyzed and classified;
- an urgent problem of enterprise-level integrated software design and implementation detected and stated;
- integrated data and metadata model suggested as a generalized approach to solve the problem formulated;
- system interfaces and supporting architecture designed for a versatile general-purpose enterprise-level IS;
- fast IS prototype and full-scale HRIMS implemented.

Research methods meeting the problem domain specific features are based on a creative synthesis of fundamental statements of finite sequence, category and semantic network theories.

In particular, a data model is introduced that provides integrated problem-oriented event-driven data and metadata dynamics and statics management of heterogeneous weak-structured problem domains in a more adequate way than previously known ones.

Another aspect of the research novelty is a system architecture solution for open, distributed, interoperable environment supporting front-end various data warehouse processing (from high-end databases to legacy IS) based on dynamic SQL with stored procedures and object-oriented CORBA, UML and business-process reengineering (BPR) technologies.

## 2. Existing Approaches

Current approaches to HRIMS are based on data model, system architecture, IT, DBMS and CASE technologies.

HRIMS software existing can be roughly divided into three major categories:

a) mainframe-based inventory control (IC) approach;

b) MRP and ERP distributed approach;

c) enterprise application integration (EAI) approach.

Let us briefly describe each of the approaches named.

### 2.1 Conservative inventory control approach

First enterprise-level solutions have been obtained, apparently, in 60-s as custom-made IS for IC based on mainframes. Examples include hierarchical (*IBM IMS*, 1968) and network (*Cullinet Software IDMS*, 1971) systems.

Some of the systems built under this approach include such advanced features as payroll data entry, storage and reporting, multimedia personal data handling (including employee photos, etc.), ordinary and sick leave data entry, storage and reporting, flexible form and report generation.

In most cases a high security level is guaranteed by explicitly enabling/disabling rights of every user for each entry form, report or query. Since systems of the kind are non-client/server ones, they lack flexibility of a distributed system and it is hard to develop and deploy applications and perform WWW data publishing.

A positive example of systems concerned is *Q Data Dynamique (Pty) Ltd. UniQue* HRIMS originally designed for use with AS/400 and later on (in 1993) adapted for PC LANs. A present-day *UniQue* implementation example is Russian subsidiary of *Coca-Cola Refreshments Co.* HR and payroll data are stored together in a uniform custom database making an integrate solution.

However, systems of the kind are not based on any particular data model and therefore they use rather primitive set of standard functions and neither possess a front-end programming and development environment nor can easily accept data from foreign sources.

### 2.2 Advanced Approaches: More Integration and Flexibility

**From MRP to EAI: Standards and Models Evolution**

In 70-s to solve the problem of on-line and analytical enterprise resource management MRP I-II software systems appeared. The MRP information systems were based on DBMS built in accordance with E.F.Codd's relational model [4] (including prototypes of *System R* [11], *IBM*, W.Kim, 1968 and *Ingres* [20] M.Stonebraker, Berkeley University, 1976-79) and respective commercial systems of *IBM DB2* and *CA Ingres*. Earlier versions of *SAP R/3* can be considered as a positive example of MRP systems, capable of more generalized resource control than outdated IC systems. MRP solutions supported dynamic SQL dialects and were aimed at small (up to 1,000 people) and medium (up to 10,000 people) corporations.

Further development of data models (such as P.Chen's ERM [3], 1980), data object manipulation languages (SQL2, 1980), industrial DBMS [15] (*Oracle, Informix, Sybase*) and integrated CASE-tools has resulted in appearance in early 90-s of client/server-based enterprise resource planning (ERP) systems providing consolidated resource management for large (more than 10,000 people) corporations.

Within adjacent field of heterogeneous system integration, parametric generalizations of methods and tools with general-purpose software development have been obtained in late 90-s by D.Calvanese [2], D.Florescu, A.Levi [6] et al. (ODBC/JDBC data integration), D.Linticum [12], H.Davis [5] and others (COM/CORBA application integration) and Y.Kambayashi (Java, ActiveX, MOM and RPC interface integration).

Two most successful examples of state-of-the-art HRIMS include *Oracle* and *Lotus Development Corporation* systems.

**Oracle Human Resources**

*Oracle Human Resources* services are based on highly scalable and reliable relational database *Oracle Universal Server*. The HRIMS features compact storage, effective retrieval of multimedia data, advanced form generator and report writer, object-oriented visual interface script language, SQL-based procedure-oriented query language, cross-platform support, WWW-ready applications development.

Personnel data can be compactly stored in a complex multimedia format including photos of the employees, their signature samples, interviews, video records, certificate color copies, etc. and retrieved in a fast way.

Multi-platform client/server support is provided for most of leading operating systems including *MS Windows NT*, *Sun Solaris*, *IBM AIX* and other *UNIX* dialects etc.

Developed and deployed database-oriented applications become Web-enabled through *Oracle Web Server*.

*Oracle Designer/2000* allows to enhance and optimize HR-oriented applications using visual interface and an SQL-based *PL\SQL* language as a basis and a visual object-oriented script language at upper level. Data exchange and database integration is possible with *Oracle Applications* group *Oracle Financials* modules which include *General Ledger*, *Payables, Receivables, Assets, Manufacturing, Project Management,* and *Purchasing*.

For example, there are certain points of integration within *Oracle Assets* which can use personal data from *Oracle Human Resources* for depreciation and tax calculations.

However, *Oracle Applications* group products are integrated loosely enough and much is still desired to build a real enterprise level solution out of them.

**Lotus Notes Solution for HRIMS**

Understanding the field importance, *Lotus Corporation*, the leader in groupware production [15], introduces a concept of a HR system [13].

The system is based on *Lotus Domino Server* and includes multimedia and non-structured data handling, mail services for HR collaborate activities, WWW-ready environment, electronic routing and approvals, streamlined scalable workflow, telephony-based solutions, scheduling for personnel training, cross-platform support, ODBC integration, advanced replication, mobile users support and high security.

*Lotus* provides a basis for a wide spectrum of HR activities including recruiting, applicant training, advertising, job offering negotiations and approvals, hiring support, automated inquiry service, personnel activities testing, assessment and training, performance management, compensation and benefits administration and HR call support.

Though *Lotus* provides a truly hi-tech groupware-oriented flexible solution for HRIMS, the database is too open (the approach is more applicable to recruiting agencies than to enterprises) and lacks some important HRIMS features such as integrated appraisal and testing as well as extra source data interconnection except ODBC.

There are, however, more versatile approaches existing which are a combination of the above mentioned ones. One of the positive examples is *Oracle InterOffice* [14].

## 3. Related Works

Papers [1,8,16-19] provide rigorous mathematics foundation and solid theoretical research background.

Lattice of flow diagrams which can be used to represent and to model data flow is discussed in [17].

Object hierarchy as a basic approach to handling objects storage and manipulation is described in [7]. Database structure can also be illustrated in detail using the semantic networks approach introduced in the paper.

Papers [1,4,9-11] deal with various database structure notations, which is necessary to be taken into consideration. Relational DBMS and weak-structured document solutions are cross-examined.

Semantic networks theory is developed in [6]. An intuitively transparent way to adequately illustrate both explicit and obscure object properties has evolved from this paper.

Modern viewpoints on object-based implementation are presented in [22,23]. Using a rigorous mathematical object foundation, the papers provide general overview of object-oriented systems theory suggesting a number of practically applicable solutions as well.

Enterprise groupware-based HRIMS solution is outlined in [24] and given an even more wide coverage in [25,26]. Recently developed advanced HRIMS overview (see Section 2) is based on thoroughly studied user and system documentation from respective vendors [21]. World recognized independent expert opinion and advise [15] are also taken into consideration. Current HRIMS status is acquired from World Wide Web [13,14].

## 4. Architecture and Interface Assumptions

According to personnel management problem domain research results, vital issues of problem-oriented HRIMS construction for integrated corporate resource management have been formulated. In accordance with problems detected, fundamental requirements for versatile enterprise-level software design and implementation have been classified. Attention has been concentrated on programming, mathematics, information and language support.

Specific features of the problem domain demand support for dynamic multilevel personnel restructuring process with multi-alternative assignments based complex estimation of employee activity. In respect of HRIMS interface the requirements set should allow:

- mandatory input fields dynamic variation;
- access rights flexible differentiation;
- continuous support of data completeness and integrity.

In architecture respect, the system should provide the following features:

- openness;
- expandability;
- flexible adapting to problem domain state;
- data and metadata correction possibility (including rollback mechanisms).

## 5. The Integrated Data and Metadata Model

### 5.1 The Data Object Model

Mathematical formalisms existing for problem domains are not fully adequate to dynamics and statics semantic peculiarities. Besides, modern methods of CASE-and-RAD design and implementation of integrated enterprise applications do not result in solutions of a wide application spectrum; the corresponding commercial HRIMS do not provide a significant degree of complex heterogeneous problem domains data usage.

According to results of research on enterprise personnel management problem domain specific features, a computational data model (DM) based on object calculus has been built. The model is a theoretical method synthesis of finite sequences, categories and semantic networks.

Date objects (DO) of the DM introduced can be represented as follows:

DO = < concept, individual, state >,

where a concept is understood as a collection of functions with the same definition area and the same value range. An individual implies an essence selected by a problem domain expert, who indicates the identifying properties. State changes simulate dynamics of problem domain individuals.

Compared to research results known as yet, the DM suggested enjoys significant advantages of more adequate dynamics and statics mapping of heterogeneous problem domains, as well as support problem-oriented integrated data management. In architecture and interface aspects the DM provides straightforward iterated design of open, distributed, interoperable HRIMS based on UML and BPR methodologies. As far as implementation is concerned, information processing from various repository types of heterogeneous enterprise problem domains is supported providing front-end data access based on event-driven procedures and dynamic SQL technologies.

The computational model suggested is based on the two-level conceptualization scheme [22], i.e. process of establishing relationship between concepts of problem domain.

Individuals h, according to the types T assigned, are united in assignment-depending collections, thus making variable domains of sort

$$H_T(I) = \{h \mid h : I \to T\},$$

that simulate problem domain dynamics.

When fixing data model individuals, uniqueness of individualization of data object d from problem domain D by means of the formula $\Phi$ is required:

$$\| Ix\ \Phi(x) \|\ i = d \Leftrightarrow \{d\} = \{d \in D \mid \|\Phi(d)\|\ i = 1\}.$$

### 5.2. The Metadata Object Model

Compression principle for the computational data object model introduced

$$C = Iy: [D]\ x : D(y(x) \leftrightarrow \Phi) = \{x : D \mid \Phi\}$$

allows to apply the model to concepts, individuals and states separately, as well as to data objects as a whole.

The computational metadata model expands traditional Codd's ER-model by a principle of compression:

$$x^{j+1}\ Iz^{j+1}: [\ldots[D]\ldots]\ \forall x^j: [\ldots[D]\ldots]\ (z^{j+1}(x^j)\ \Phi^j),$$ where

$z^{j+1}, x^{j+1}$ – metadata predicate characters in relation to level j,

$x^j$ - individual of level j,

$\Phi^j$ - data object definition language construction of level j.

The suggested comprehensive model of objects of the data, metadata and states is characterized by structuredness, scalability, aggregation, metadata encapsulation, hierarchy structure and visualization.

Expandability, adequacy, neutrality and semantic correctness of the formalism introduced provide problem-oriented software design with adequacy maintenance at all stages of implementation.

Semantics of computational model of objects of the data, metadata and states is adequately and uniformly formalized by means of typed λ-calculus, combinatory logic, and semantic network-based scenario description.

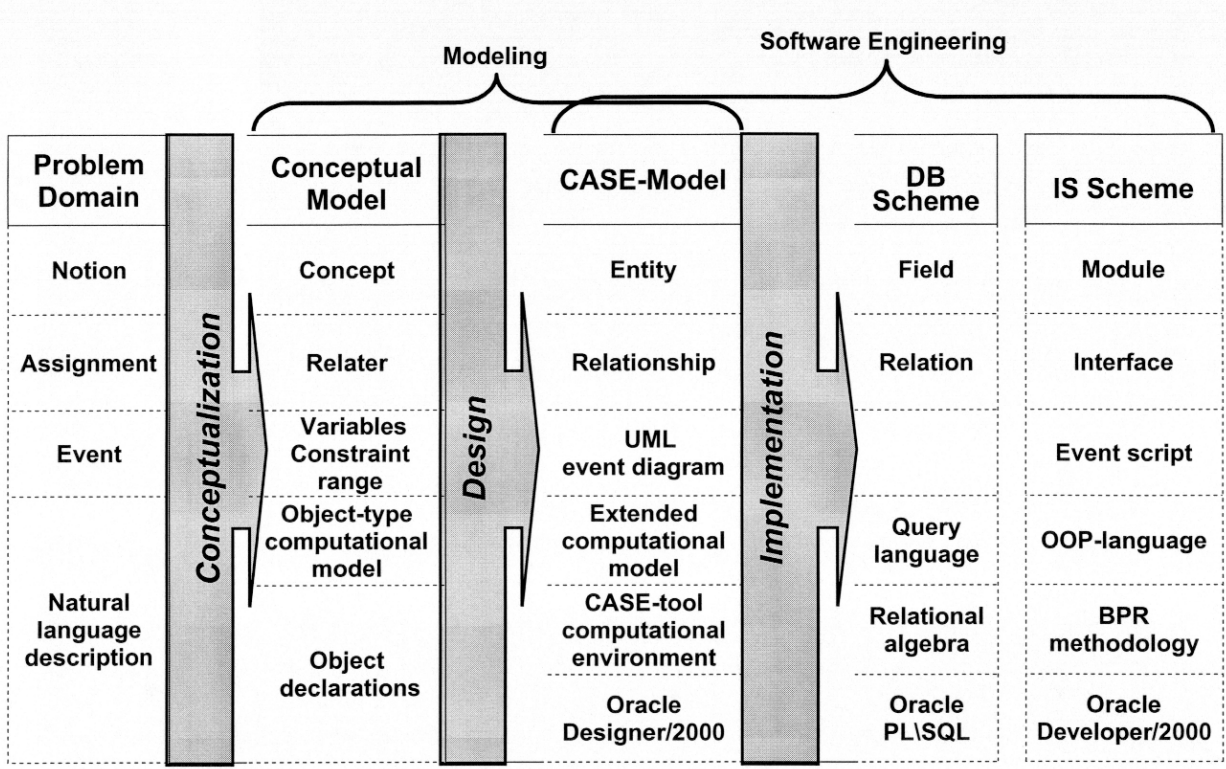

**Figure 1 Generalized implementation scheme for enterprise information systems**

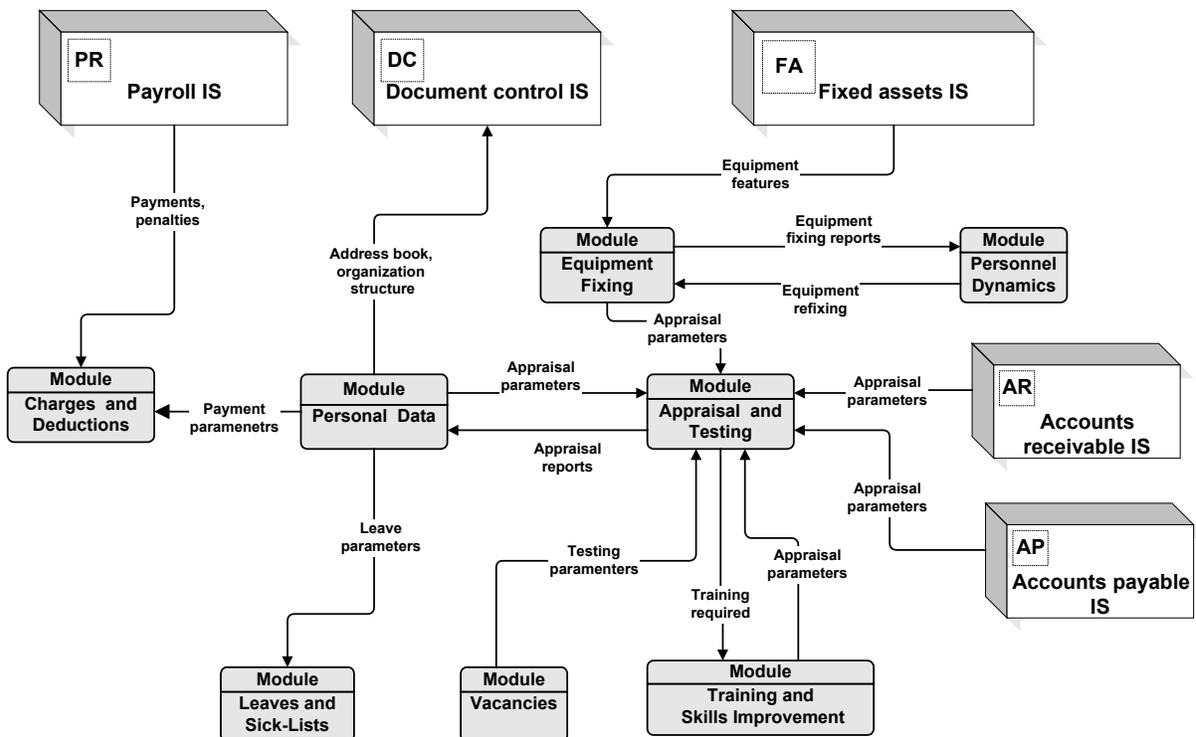

Worksł 5

**Figure 2 Implemented enterprise HRIMS structure**

## 5.3 The Integrated Model Application to HRIMS

**Personnel Appraisal Model**

Judging by specific features detected during HRIMS problem domain analysis, an appraisal modeling formalism is introduced. In the below example most essential employee evaluation parameters:

- hierarchy-based corporation organizational structure;
- employee working functions;
- vacant positions currently available;
- enrolled employee amount

are combined.

The last three parameters can be detailed from the corporation level down to its smallest structural unit, while the first parameter is a global one.

On the basis of the introduced multi-parameter functional

$F = F ((v), (e), \ldots) (s) (p)$, where

s - assignment taking into account working functions of corporation personnel;

p - assignment taking into account corporation organizational structure;

v - amount of corporation vacancies;

e - amount of corporation regular employees,

semantic object model and formal generalized procedure are constructed for comprehensive appraisal based on the F functional evaluation mapping value calculation.

**Generalized Component-Integration Algorithm**

According to the enterprise-level software design and implementation scheme, a generalized algorithm of new components integration into an existing information system structure is suggested.

The algorithm is based on the semantically preferred data objects analysis and provides consistency and integrity of extendable data object models as well as possibility of iterated information system design through by business model reengineering.

The algorithm unifies object-based heterogeneous management information system integration process.

An important feature of the generalized component-integration algorithm is its semantic orientation. In terms of human resources information management system it implies organizational structure dependency.

The author has performed research of corporation organizational structure semantics earlier. Research results are presented in [25,26], where they are discussed in a more detailed way.

## 6. UniQue: a Full-Scale Integrated Enterprise HRIMS Implementation

### 6.1 Generalized Implementation Scheme and its Customization for HRIMS

The multilevel integrated information system (IS) design and implementation scheme providing component-wise development of the integrated open extendable corporate enterprise-level software with continuous control of adequacy and integrity (see fig.1) is suggested. During design process, IS specification is transformed from problem domain concepts to data model essences and further through CASE-tools to DBMS scheme with *PL\SQL* as data object manipulation language to target IS description with appropriate architectural and interface components.

In accordance with problem domain specific features analysis results, computational data model and generalized scheme of IS development have been adapted to satisfy the required personnel management conditions. The problem-domain oriented IS design scheme includes five stages:

- corporation board of directors formulates objectives, measures and plans on restructuring which are mapped in formal business rules of IS computational model;
- experts in personnel and information create the specified structural and functional conceptual corporation business model in a form of object map;
- system analysts make OLAP-research of corporate business model variants for various development scenarios;
- database and IS developers formalize business logic of the architecture and interfaces using object-oriented script language, language, which is assembled in UML-data model by means of CASE-synthesis methodology;
- database, local area network and security managers implement and support target IS and database schemes.

## 6.2 Designing Problem-Oriented Interface and Supporting Event-Driven Architecture

According to detailed enterprise-level software design sequence, a generalized heterogeneous repository processing scheme is introduced that allows users to interact with distributed database in a certain state depending on dynamically activated (i.e., assigned) scripts. Thus, the scripts in a form of database connection profiles and stored object-oriented program language procedures are initiated depending on user-triggered events. Scripts provide transparent and intellectual client/server front-end user-to database connection. Dynamically varied database access profiles provide high fault tolerance and data security both for ordinary and privileged system users in the heterogeneous environment. The profiles are implemented using CORBA technology as an intellectual media between end user and the heterogeneous data warehouses.

Depending on semantic-oriented corporation organizational structure, which defines HRIMS user position in the corporation hierarchy, a certain database connection and access level profile is assigned dynamically. The profile is valid only until the end of a data exchange session. According to the corporation hierarchy, user gets access to data under one of the basic scenario profiles ranging from corporation president to a department employee. Access is granted not only to data, but also to metadata (i.e., data object dimensions, integrity constraints, access rights, appraisal parameters, corporation structure etc.).

Administrative users have extended access to metadata.

Thus, under the model introduced data and metadata objects are manipulated uniformly. This makes system interface a problem-oriented, straightforward and uniform one and significantly increases system performance and user-friendliness.

## 6.3 Implementation Description

The introduced design methodology has been practically approved during *UniQue* HRIMS implementation at *ITERA* international corporation. The enterprise personnel management software consists of eight components (see fig.2). *Personal Data* component is intended for storage and processing of employee biography data. The subsystem, connected to the above one, named *Personnel Dynamics* allows to control dynamics of employees enrollment, transfer, dismissal and re-enrollment events. The adjacent software module *Charges and Deductions* provides registration of salaries, bonuses and other kinds of payments, as well as material penalties. The central *Appraisal and testing* component supports comprehensive employee labor activity estimation based on individual, psychological, professional and other kinds of tests, as well as on adjacent *UniQue* modules and third-party software data. *Vacancies* component supervises personnel selection according to given criteria. *Leaves and Sick-Lists* subsystem accounts employees working hours and supports multi-type leaves. *Training and Skills Improvement* component implements training policy judging by appraisal results and tracks training service payments. Finally, *Equipment Fixing* subsystem provides registration for accountable persons and major corporation resources used by them.

All of the HRIMS components are captured by unified interface and integrated into enterprise IS environment of the *Oracle Applications* financial and commodity management systems and *Oracle InterOffice* document management system.

From the system architecture viewpoint the integrated *UniQue* HRIMS provides certain level of data input, correction, analysis and output (from president down to chief of a department) depending on front-end position (i.e., assignment) in personnel hierarchy. Problem-oriented form designer, report generator, on-line documentation and administration tools are used as interactive interface facilities. *UniQue* database supports the integrated storage for data (for on-line user access) and for metadata (data object dimensions, integrity constraints and other business process parameters).

During the personnel management corporate-level IS design process problem domain data model specification (represented as semantic network fragments) has been transformed into use-case UML diagrams, then, by means of *Oracle Developer/2000* integrated CASE-tool - into ER-diagrams and, finally, into the attributes of target IS and databases.

On the basis of the information model developed, architecture-and-interface solution for integrated personnel management software has been designed; details of database processing for various system user and administrator classes have been considered.

Software implementation has been divided into two stages: 1) fast prototype created with an SQL-based query language, supporting triggers and stored procedure mechanisms using *PowerScript* script object-oriented language and 2) full scale and capacity software implementation based on the *Oracle* integrated information system development tools platform.

To prove adequacy of the computational data and metadata model developed and component integration algorithm suggested, a fast software prototype has been designed on the basis of generalized architecture scheme and supporting interfaces.

*Sybase S-Designor/PowerBuilder* has been chosen as CASE- and RAD-toolkit for implementation environment as a result of carried out comparative analysis.

According to prototype approbation results full-scale object-oriented software has been implemented and subsequently adapted for personnel management application development.

To provide required levels of industrial scalability and fault tolerance, judging by the results of CASE-and-RAD tools multi-criteria comparative analysis *Oracle Developer/2000* toolkit has been chosen as an integrated solution supporting methodologies of universal modeling (UML) and business processes reengineering (BPR) methodologies.

Target IS implemented consists of eight components using a set of Oracle tools. All of the components are implemented according to technical specifications designed by the author personally, and amount to more than 1000 source text pages. 150 high complexity bilingual screen forms and reports, as well as about 30-page source text size *Equipment fixing* component have been also created by the author. According to specification requirements developed by the author together with *ITERA* corporation personnel service the software implemented had been significantly improved. In particular, procedures of accounting salaries and vacation bonus have been coded.

The full-scale implementation is based on the hardware platform of an IBM RS/6000 two-server high availability cluster running under AIX operating system with 512 Mbytes RAM and 60 Gbytes SSA disk array with increased reliability and fault tolerance level similar to RAID of level 5.

Average response time of the *UniQue* enterprise-scale IS makes about 6,5 seconds.

The information system has been implemented in a conglomerate international corporation and has passed a two-year experimental check.

As a result of software implementation designed on the basis of the model introduced, implementation terms and cost and cost compared to existing commercial software of the kind are considerably reduced while the functional set is extended.

In the opinion of users, the software implemented features high degrees of openness, expandability, flexibility, reliability, ergonomics and ease of mastering.

Thanks to problem-oriented interface, primary data entry speed exceeds that of commercial software of the kind by the average of 20% and amounts to about 150 seconds per employee data entry. Access levels differentiation allows to considerably result risk of information distortion or loss.

## 7. Results

1. Computational data model has been introduced providing integrated manipulation of data and metadata objects especially under the conditions of rapidly changing heterogeneous problem domains. The model is an alloy of methods of finite sequences, category theory and semantic networks.

2. An original generalized scheme of "straightforward" enterprise IS design and implementation has been proposed on the basis of formal data and metadata model. The scheme includes a for new components integration into an enterprise-scale IS that provides adequacy, consistency and data integrity.

3. According to the above mentioned scheme and algorithm, a generalized IS interface has been designed based on an open and extendable architecture.

4. To solve a complex applied task of enterprise resource management, fast event-driven prototype software has been developed on the basis of generalized interface and architecture of structural and logical UML scheme.

5. Using the prototype approbation results, a full-scale object-oriented software has been designed and applied for a versatile enterprise-level implementation.

6. The full-scale enterprise-level software has been customized for corporate personnel management and implemented at a corporation with more than 1000 people staff.

## 8. Conclusion

Results for enterprise-level personnel management solution implemented have proved significant decrease in terms and costs of implementation as well as growth of portability, expandability, scalability and ergonomics levels in comparison with existing commercial software of the kind. Iterated multilevel software design is based on formal model synthesizing object-oriented methods of data (data objects) and knowledge (metadata objects) management. Industrial implementation of the enterprise-level HRIMS software has been carried using integrated CASE- and RAD-toolkits. Practical implementation experience has proved importance, urgency, originality and efficiency of the approach as a whole as well as of its separate stages and solutions.

Theoretical and practical statements outlined in the paper have been approved by *UniQue* enterprise-level HRIMS software successful implementation at *ITERA* international energy corporation.